\documentstyle[12pt,aps]{revtex}
{
\begin{document}
\draft

\title
{NUMERICAL STUDY OF THE PAIRING CORRELATION OF THE $t-J$ TYPE MODELS}

\author{ C.T. Shih$^1$, Y.C. Chen$^2$, and T.K. Lee$^3$}
\address{
$^{1}$National Center for High-Performance Computing, Hsinchu, Taiwan\\
$^{2}$Dept. of Physics, Tunghai Univ., Taichung, Taiwan\\
$^{3}$
Inst. of Physics, Academia Sinica, Nankang, Taipei, Taiwan;
and National Center for Theoretical Sciences, Hsinchu, Taiwan
}

\date{\today}
\maketitle
\begin{abstract}

We reported that the pair-pair correlation function of the two-dimensional
$t-J$ model does not have long-range d-wave superconducting correlations in
the interesting parameter range of $J/t \leq 0.5$. The power-Lanczos method
is used under the assumption of monotonic behavior. This assumption has
been well checked in the two-dimensional $t-J$ and attractive Hubbard model.
Here we re-examine this criterion of monotonic behavior of the pairing
correlation function for the one-dimensional and two-leg $t-J$ ladder where
other accurate numerical results are available.
The method seems to be working well.

\end{abstract}

It has been shown that
several essential physical properties of the high temperature
superconducting (HTSC) cuprates can be described by the
two-dimensional (2D) $t-J$ model \cite{anderson87,zhang88},
such as the single hole dispersion relation \cite{lee96}, non-Fermi
liquid behavior \cite{eder97}, and phase separation
\cite{shih98a}. One of the critical issues is whether
the model gives large enough pairing correlation to quantitatively
explain the high $T_c$.
We reported part of the numerical
results on this issue in Ref.\cite{shih98b} which concludes
that the 2D $t-J$ is {\it not} superconducting in the physical
interested region of electron density $n_e$ and coupling
constant $J/t$ from the studies on the pair-pair
correlation function and the two-hole binding energy\cite{Chernyshev98}.

The criterion to evaluate the pairing correlation of
the 2D $t-J$ model used in Ref.\cite{shih98b,heeb94}
is to apply the power-Lanczos(PL)
method \cite{yctk95} on the variational trial wave functions.
We denote the initial trial wave function as $\mid PL0\rangle$
and the one improved by one Lanczos iteration as $\mid PL1\rangle$.
We calculate the change of the pairing correlations of
$\mid PL1\rangle$ and $\mid PL0\rangle$ for various variational
parameters. If we assume the behavior of the long-range pairing
correlations which is monotonic through the Lanczos or power iteration,
then the one whose pairing correlation is least
changed gives the correct results of the ground state.
It is extremely difficult to prove this criterion by
 the present computing power in 2D, we could only provide more
examples to support it. In a previous paper we have checked the
validity of the criterion of the 2D attractive Hubbard model\cite{shih98b}.
Here we will show the results of the one-dimensional(1D) $t-J$ model
and 2-leg $t-J$ ladders.

The $t-J$ model is:
\begin{equation}
H=-t\sum_{<i,j>\sigma} (\tilde{c}^\dagger_{i\sigma}\tilde{c}_{j\sigma}+h.c.)
  + J\sum_{<i,j>}({\bf S}_i\cdot{\bf S}_j-{1\over4} n_in_j),
\label{e:1dtj}
\end{equation}
where $\tilde{c}^\dagger_{i\sigma} = c^\dagger_{i\sigma}(1-n_{i-\sigma})$, and
$\langle i,j\rangle$ denotes the nearest neighbors $i$ and $j$.

For the 1D case, we study the pairing correlation function in the spin-gap
region of the phase diagram, i.e., large $J(>2)$ and low electron density
\cite{spingap1,spingap2}.
Here we use the correlated spin-singlet pair(CSP) wave function as
$\mid PL0\rangle$. The CSP wave function is defined as

\begin{equation}
\left|CSP\right>= P_d \Pi_{i>j}\left[\frac{L}{\pi}sin\left(\frac{\pi}{L}
(r_i-r_j)\right)\right]^\nu[\sum_{n=1}^{\infty} h^{n-1}
b_{n}^{\dagger}]^{N_{e}/2} \left|0\right>,
\end{equation}

where  $h=2t/J$. $N_{e}$ is the total number of electrons,
$P_d$ is the projection
operator that forbids two particles occupying the same site and the operator
$b_{n}^{\dagger}= \sum_{i} c_{i\uparrow}^{\dagger} c_{i+n\downarrow}^{\dagger} -
c_{i\downarrow}^{\dagger} c_{i+n\uparrow}^{\dagger}$.
The variational parameter $\nu$
control the long-range correlations.
In Fig. 1 we show the pairing correlation function of the
$\mid PL0\rangle$ and $\mid PL1\rangle$ with $J/t=2.8$.
In order to check the ansatz
of monotonic behavior, we choose a small lattice of 24 sites with 4 electrons
which can be exactly diagonalized by Lanczos iteration. As it clearly
show that the one with $\nu=0.1$ has the stablest long-range correlation
which is also very close to the value of the exact ground state. For smaller
$\nu$ the long-range correlation decrease while the larger ones increase.
The results for this 1D case are consistent with our ansatz.

For the ladders discussed here, the leg-leg and rung-rung $t$ and $J$ are
assumed to be identical.
In the variational level of the $2\times L_x$ $t-J$ ladder,
the trial wave function used here is\cite{gros88,zhang96}:
\begin{equation}
\mid \Phi(\Delta,\nu) \rangle = \prod_{i<j}\{\frac{L}{\pi}sin(\frac{\pi}{L}
(x_i-x_j)\}^{\nu}\cdot P_d \prod_k (\tilde{u}_k +
\tilde{v}_k c^\dagger_{k,\uparrow}
c^\dagger_{-k,\downarrow})\mid 0 \rangle
\label{e:TWRVB}
\end{equation}
with
$\tilde{v}_k/\tilde{u}_k =
\Delta_k/(\epsilon_k + \sqrt{\epsilon_k^2
+\Delta_k^2})$,
 $\Delta_k = \Delta(cosk_x - 2cosk_y)$ and
$\epsilon_k = (2cosk_x + cosk_y) - \mu$.
 $\Delta$ is the d-wave superconducting order parameter
and $\mu$ is the chemical potential. The operator $P_d$ enforces the
constraint of no double occupancy. We take $t=1$ in this paper.
In the 2-leg ladder case, the rung-rung pairing correlation
$P(R)=\langle \Delta_i^{\dagger} \Delta_{i+R}\rangle$
is measured, where
$\Delta_j=(c_{j,2;\downarrow}c_{j,1;\uparrow} - c_{j,2;\uparrow}
c_{j,1;\downarrow})$.

For the 1D and ladder cases, because the correlation functions
decay exponentially or in power law with respect to the distance
$R$, rather than being flat plateaus in two dimensions,
it is difficult to define the ``stable''
correlation of the trial wave functions with different parameters
$\Delta$ and $\nu$. Thus we define the deviation $D(\Delta,\nu)$
to quantify the difference between the $\mid PL0\rangle$ and
$\mid PL1\rangle$ wave functions. The wave function with minimal $D$
is the one having stable correlation. It is defined as
\begin{equation}
D=\frac1N\sum_{R\ge2} \{(P_1(R)-P_0(R))/P_0(R)\}^2
\label{e:diff}
\end{equation}
where $P_0(R)$ and $P_1(R)$ are the rung-rung pairing correlations,
respectively. Note that this is not the unique way to define the
deviation. We tried some other reasonable definitions of deviation
and the results are consistent.
Fig.2 shows the wave function with minimal deviation
of 12 electrons (6$\uparrow$ and 6$\downarrow$)
in the $2\times 8$ ladder and $J/t=1$. The wave
function with minimal deviation is $\Phi(\Delta=0.12,\nu=0.6)$.
Fig.3 shows $P_0(R)$ and $P_1(R)$ of the trial
wave functions and the exact results of the system. The ``sandwich''
behavior of the long range part (R=3, 4) is clear.

Since $2\times 8$ is too small to see the long rang behavior of
the pairing correlation, we did the similar analysis on $2\times 30$
system, which the DMRG results are available for comparison\cite{hayward95}.
We tried $24\uparrow$ and $24\downarrow$ electrons in $2\times 30$ ladder with
$J/t=1.0$. The best one (minimal deviation) is $\Delta=0.16$ and
$\nu=0.3$, while the trial wave function with lowest energy
is $\Delta=0.24$ and $\nu=0.1$, whose pairing correlation
is evidently overestimated.
The pairing correlations obtained with three different
sets of parameters are compared with  the DMRG
data from \cite{hayward95}
in Fig.4. It can be seen that the slopes in the log-log
plot, that is, the exponents of the power law decay, of the
 trial wave function (-0.835) with minimal $D$ and the DMRG data
(-0.840) are very close. The differences of amplitudes
and slopes of the long tails
between these two may come from the different boundary conditions of
DMRG (open boundary condition, OBC) and power-Lanczos method
(periodic boundary condition, PBC). In fact, we have seen similar difference
between OBC and PBC by solving
the $2\times8$ ladder with 12 electrons and $J/t=1$
exactly, which is shown in the inset of Fig.4.

In summary, from the study of 1D and 2-leg ladder $t-J$ models,
there are more evidences to support the ``least deviation'' criterion
used to determine the superconductivity of the 2D $t-J$ model.
It seems that the pure 2D
$t-J$ model doesn't give enough amplitude of pairing
correlation in the physically interested parameter region.
Since there are a number of evidences that the 2D t-J model
is a fairly good model for HTSC, it seems the model should not
be excluded just due to the negative result with respect to
the d-wave pairing correlation. It is possible that we can
keep the good properties of the 2D model and enhance the
superconductivity by some reasonable modification, for example,
adding the interlayer hopping term\cite{bilayer}.

The authors like to thank Poilblanc and Noack for providing us the
DMRG data. This work is partially supported by the National Science Council of
Republic of China, Grant Nos. 89-2112-M-001-050 and 88-2112-M-029-002.
Part of computations were performed at the National Center
for High-Performance Computing in Taiwan. We are grateful for
their support.


\newpage
\noindent {\bf Figure Caption:}

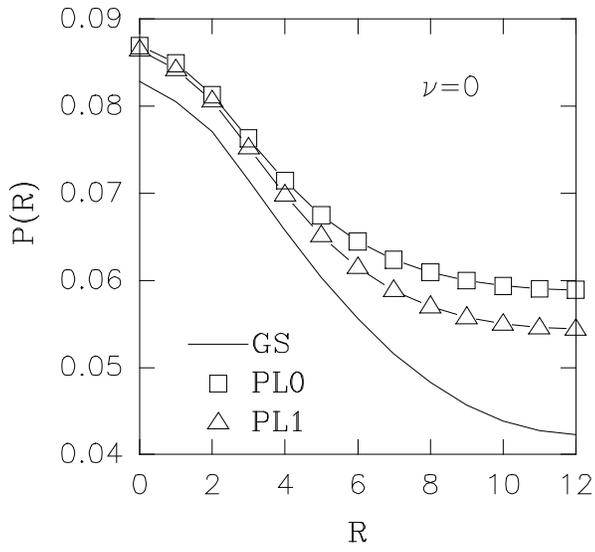
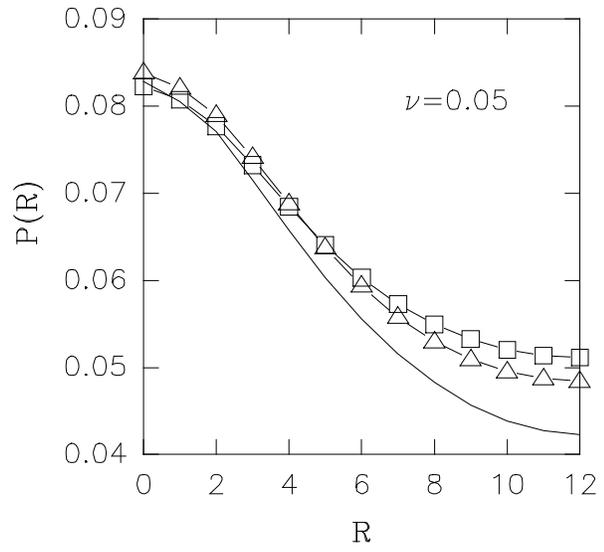
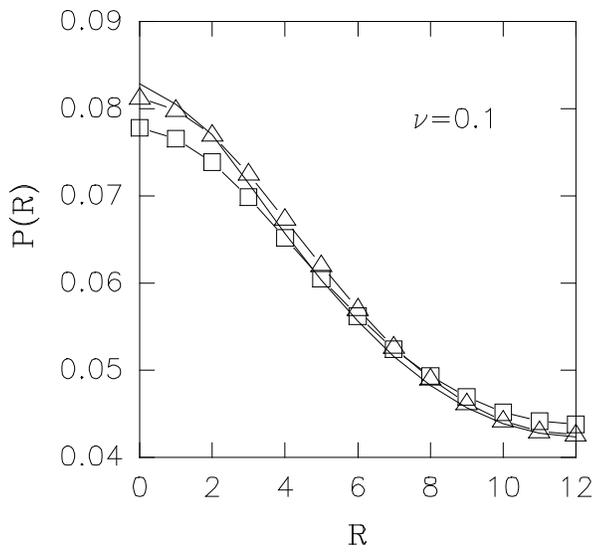
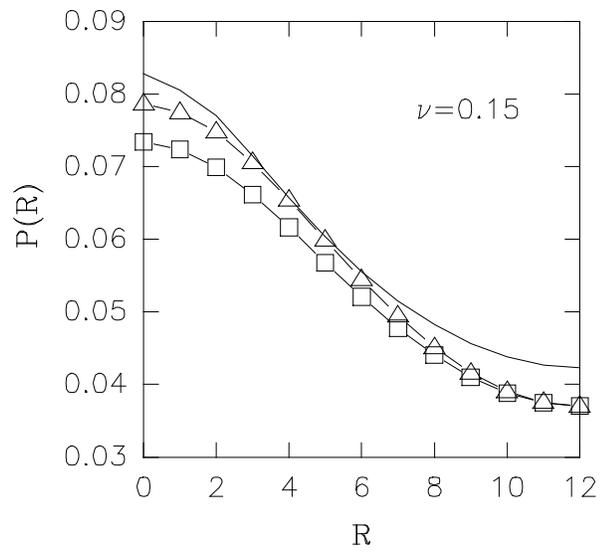
\noindent Fig.1: Pairing correlation function $P(R)$ for various
trial wave function(PL0) and their PL1 results, together with the exact
data(GS) for $J/t=2.8$ of 4 electrons in a 24-site chain.

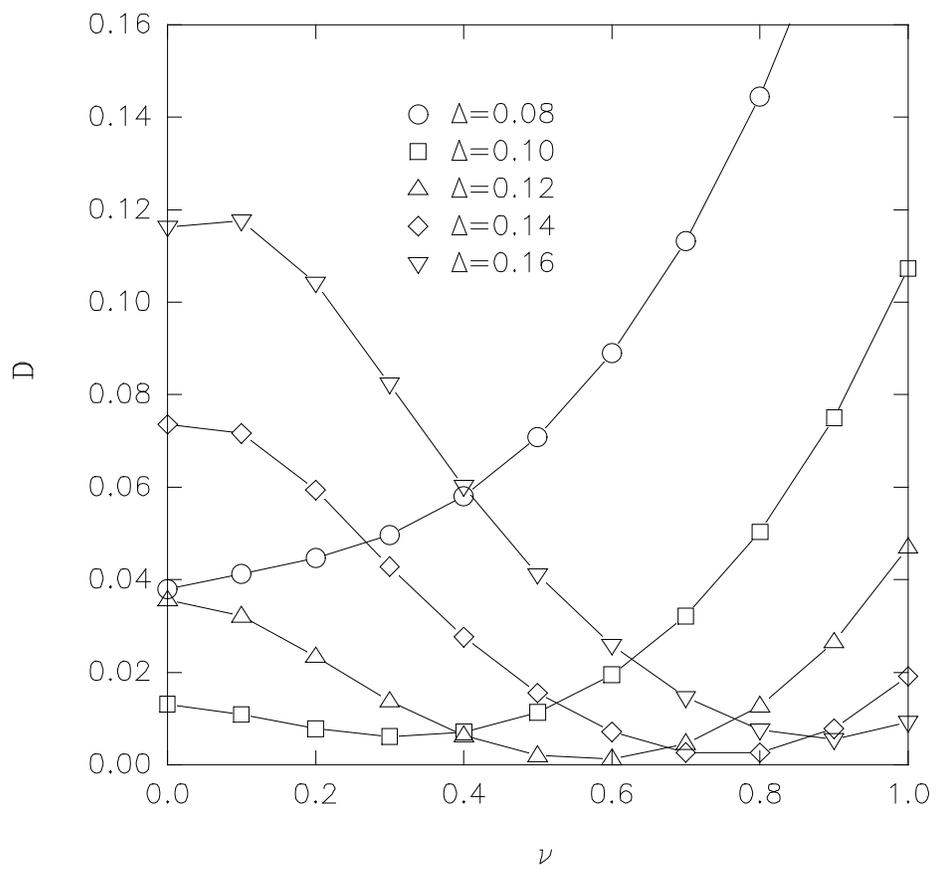
\noindent Fig.2: Deviation $D(\Delta,\nu)$ for  a $2\times 8$ t-J
ladder, $J/t=1$.

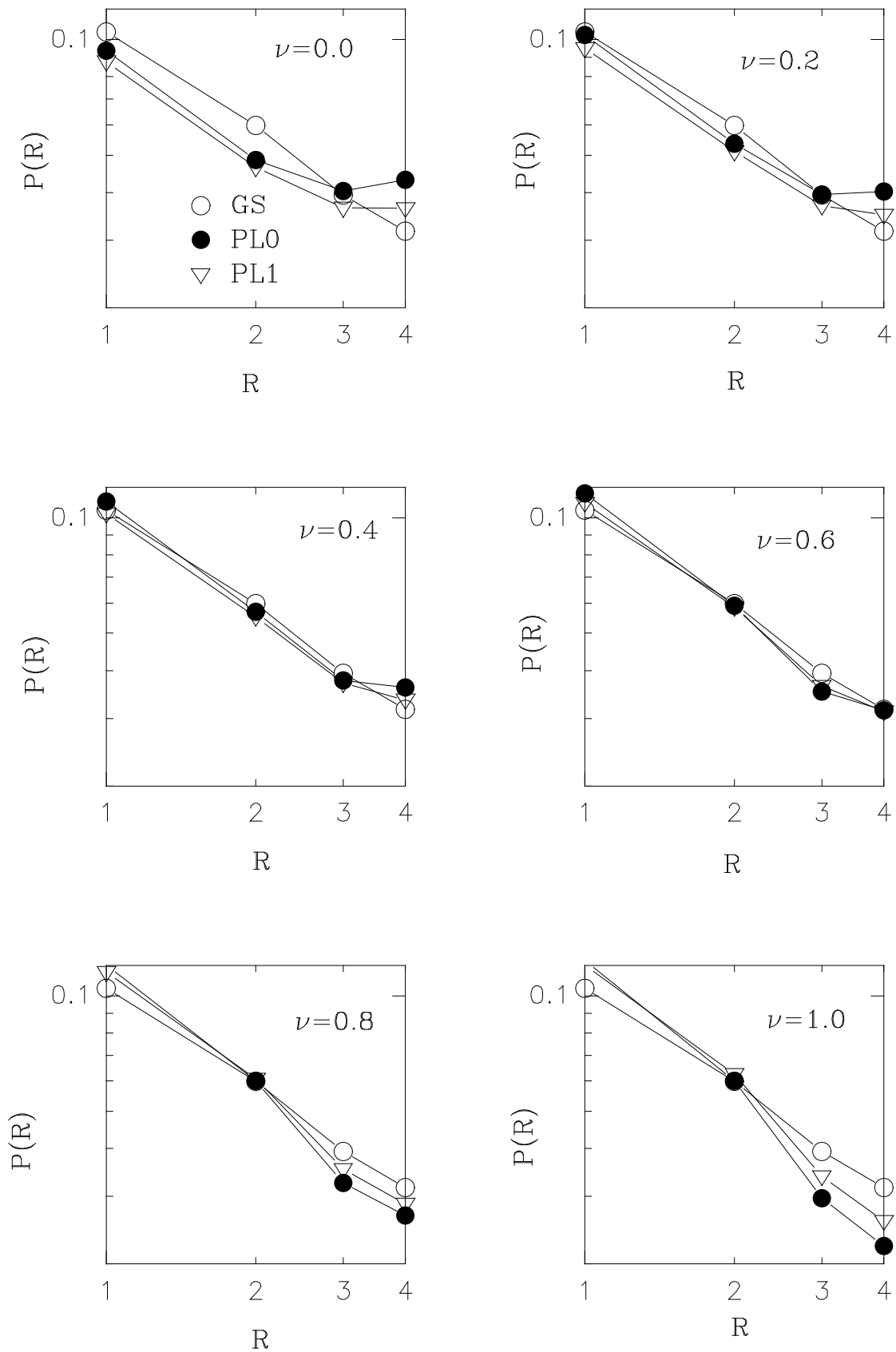
\noindent Fig.3: Pairing correlation function $P(R)$ for
trial wave functions with $\Delta=0.12$ and various
$\nu$ and their PL1 results, together with the exact
data for $J/t=1$ of 12 electrons in  a $2\times 8$ ladder.

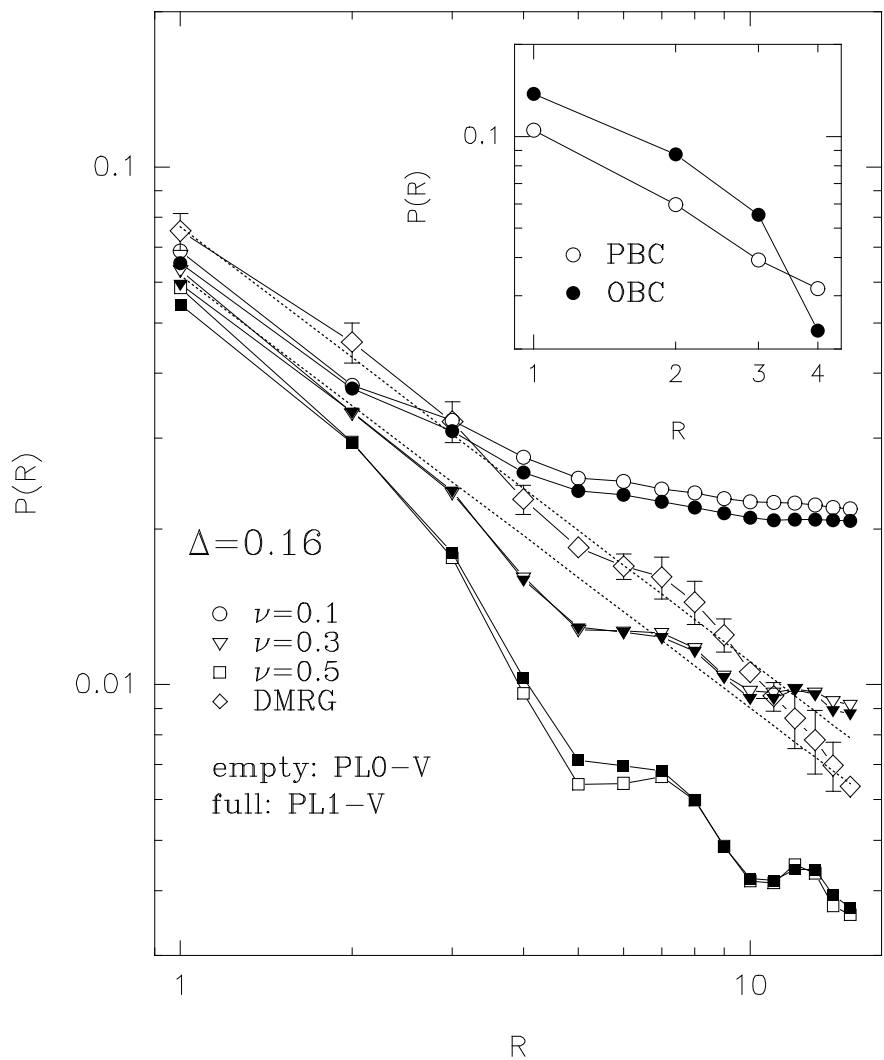
\noindent Fig.4: Similar to Fig.2, but for 48 electrons
in  a $2\times 30$ ladder. The dot lines shows the power law fitting.
The inset shows the $2\times 8$ exact results
for the different boundary conditions.


}

\begin{thebibliography}{99}

\bibitem{anderson87}P. W. Anderson, Science {\bf 235},
1196 (1987).

\bibitem{zhang88}F. C. Zhang, and T. M. Rice, Phys. Rev. B{\bf 37},
3759 (1988).

\bibitem{lee96}T.K. Lee and C.T. Shih, Phys. Rev. B{\bf 55}, 5983 (1996).

\bibitem{eder97}R. Eder, Y.C. Chen, H.Q. Lin, Y. Ohta, C.T. Shih, and T.K.
Lee, Phys. Rev. B{\bf 55}, 12313 (1997).

\bibitem{shih98a}C.T. Shih, Y.C. Chen, and T.K. Lee, Phys. Rev.
B{\bf 57}, 627 (1998).

\bibitem{shih98b}C.T. Shih, Y.C. Chen, H.Q. Lin, and T.K. Lee, Phys.
Rev. Lett {\bf 81}, 1294 (1998), and references there in.

\bibitem{Chernyshev98}A. L. Chernyshev, P. W. Leung, and R. J. Gooding,
Phys. Rev. B{\bf 58}, 13594 (1998).

\bibitem{heeb94}E. S. Heeb and T. M. Rice, Europhys. Lett {\bf 27},
673 (1994).

\bibitem{yctk95}Y. C. Chen and T. K. Lee, Phys. Rev. B{\bf 51}, 6723 (1995).

\bibitem{spingap1}Y. C. Chen and T. K. Lee, Phys. Rev. B{\bf 47}, 11548 (1993).
\bibitem{spingap2}Y. C. Chen and T. K. Lee, Phys. Rev. B{\bf 54}, 9062 (1996).

\bibitem{gros88} C. Gros,  Phys. Rev. B{\bf 38}, 931 (1988).

\bibitem{zhang96} M. Sigrist, T.M. Rice, and F.C. Zhang, Phys. Rev.
B{\bf 49}, 12058 (1994).

\bibitem{hayward95}C.A. Hayward, D. Poilblanc, R.M. Noack,
D.J. Scalapino, and W. Hanke, Phys. Rev. Lett. {\bf 75}, 926 (1995).

\bibitem{belinicher96}V.I. Belinicher, A.L. Chernyshev, and
V.A. Shubin, Phys. Rev. B{\bf 53}, 335 (1996).

\bibitem{bilayer}
S. Chakravarty et al., Science {\bf 261}, 337 (1993);
P. W. Anderson, Science {\bf 268}, 1154 (1995).

\end{thebibliography}
\end{document}